\documentclass[journal]{IEEEtran}

\ifCLASSOPTIONcompsoc
  \usepackage[nocompress]{cite}
\else
  \usepackage{cite}
\fi

\usepackage{hyperref}
\hypersetup{
    colorlinks=true,
    linkcolor=blue,
    filecolor=magenta,      
    urlcolor=blue,
}
\usepackage{graphicx}
\usepackage{lipsum}

\begin{document}

\title{Predicting Brazilian court decisions}

\author{Andr\'{e} Lage-Freitas, 
        H\'{e}ctor Allende-Cid,
        Orivaldo Santana,
        and~L\'{i}via de Oliveira-Lage
\thanks{Andr\'{e} Lage-Freitas is with Universidade Federal de Alagoas.}
\thanks{H\'{e}ctor Allende-Cid is with Pontificia Universidad Católica de Valparaíso.}
\thanks{Orivaldo Santana is with Universidade Federal do Rio Grande do Norte.}
\thanks{L\'{i}via de Oliveira-Lage is a Prosecutor at State’s Attorney’s Office at Procuradoria Geral do Estado de Alagoas.}
\thanks{April 20th, 2019. Authorship: A.L.F. proposed the research problem, supervised this work, and executed the experiments.
H.A.C. proposed and developed NLP and machine learning algorithms.
O.S. verified the methodology.
L.O.L. supervised and reviewed the Law aspects of this work. 
All authors discussed the results and contributed to the final manuscript. The authors declare that have no competing interests and that received no funding for this study.}
\\
\smallskip
\footnotesize \texttt{lage@laccan.ufal.br, hector.allende@pucv.cl, orivaldo.santana@ect.ufrn.br, livia.oliveira@pge.al.gov.br}
}

\maketitle

\begin{abstract}
Predicting case outcomes is useful but still an extremely hard task for attorneys and other Law professionals. It is not easy to search case information to extract valuable
information as this requires dealing with huge data sets and their complexity. For instance, the complexity of Brazil legal system along with the high litigation rates makes this problem
even harder.

This paper introduces an approach for predicting Brazilian court decisions which is also able to predict whether the decision will be unanimous.
We developed a working prototype which performs $79\%$ of accuracy (F1-score) on a data set composed of $4,043$ cases from a Brazilian court. To our knowledge, this is the first study to forecast judge decisions in Brazil.
\end{abstract}

\IEEEpeerreviewmaketitle

\section{Introduction}

Since Code of Hammurabi\footnote{https://en.wikipedia.org/wiki/Code\_of\_Hammurabi}, we have been trying to improve legal certainty in human relationships by making public the law and the rulings of courts.
In addition to publicizing the laws, legal systems usually provide further support to legal certainty through judicial decisions. These decisions are useful not only for judging specific situations, but also to influence society behavior by exposing the legal consequences of our actions. Thereby, predicting legal decisions is fundamental to understand the legal consequences of our actions as well as for supporting law professionals to improve the quality of their work. 

In Brazil for example, lower court judges decisions 
might be appealed to Brazilian courts (\emph{Tribiunais de Justi\c{c}a}) to be reviewed by second instance court judges. 
In an appellate court, judges decide together upon a case and their decisions are compiled in Agreement reports named  \emph{Ac\'ord\~aos}. Similar to lower court decisions, Ac\'ord\~aos include Report, \emph{Fundamentos}, Votes\footnote{C.f. Brazilian Law: \emph{Art. 489, Lei 13105/15}.}, and further metadata such as judgment date, attorneys, judges, etc. These Agreements documents are very useful for understanding jurisprudence thus guiding lawyers and court members about the decisions. For instance, attorneys often use these documents to prepare cases while judges should take them into account -- or even use them as guidelines -- for next decisions.

In order to understand Ac\'ord\~ao decisions, one has to read the subject at the summary, read the decision Report, how each judge voted in this case (Votes), and the final decision which can be unanimous or not. Moreover, each Ac\'ord\~ao might have more than one decision -- regarding one or more appealed case claims -- which can increase the Ac\'ord\~ao complexity. This problem becomes harder as there usually are hundreds -- and sometimes thousands -- of Ac\'ord\~aos related to the case on which a Law professional is working. 

A very common and extremely important task for Law professionals is to speculate how a specific court would decide given the ideas and the facts which compose the case. For example, this is useful for preparing and tuning a case to have a favourable decision. Hence, attorneys can rely on substantial assumptions on how judges will decide based on their arguments. 
Although this information can be found in public Ac\'ord\~aos, the myriads of available documents make this task very complex and error prone, even for experiment lawyers. 

In addition to Brazil, several other legal systems in the world share the very same problem of predicting legal decisions. 
The challenge is hence generalized as \emph{how to predict legal decisions with a satisfactory level of accuracy} to support the work of attorneys, judges, and other professionals such as counters and real state offices. By satisfactory, we mean that the quality of the prediction in terms of accuracy should be better -- or even higher -- than Law experts. 

Nevertheless, it still is very hard to perform any legal decision prediction with satisfactory accuracy, even though computers have been used for such challenge for decades~\cite{Loevinger1963}. 
For instance, Ashley and Br\"{u}ninghaus~\cite{Ashley2009} propose a method for classifying and predicting cases which is able to meet $91.8$\% of accuracy, however the evaluation takes only into account a small data set (146 cases).
Katz et al.~\cite{Katz2017} uses historical data for predicting USA Supreme Court decisions by classifying decisions in two and three classes and by presenting judge profiles. That approach reaches $70.2$\% of accuracy for predicting case decisions and is assessed by on a data set with $28,000$ cases. Also using data from the USA Supreme Court, Ruger et al.~\cite{Ruger2004} exposes how the prediction of Law experts performs in comparison to a trained statistical model for different Law fields by using less than 200 cases.
In~\cite{Aletras2016}, Aletras et al. uses Support-vector Machine (SVM) for predicting if cases violate Articles 3, 6, and 8 of European Convention on Human Rights of Human Rights. Their results acheived $78$\% of accuracy performed on 584  European Court cases separated by subjects.

Other related work takes advantage of machine learning techniques to support further legal tasks.
In~\cite{Long2018}, the authors propose a framework for automatically judging legal decisions by using Attention neural network models. They applied the approach for divorce decisions in China.
In~\cite{Shulayeva2017}, Shulayeva et al. separates legal principles from case facts on legal documents by using a Naive Bayseian Multimodal classifier. 
In~\cite{Elnaggar2018}, the authors proposes to use transfer learning to recognize the same words which have different meanings in different contexts, i. e., name-entity linking task.
In~\cite{Barros2018}, the authors uses Bayesian networks to classify legal decisions from a Brazilian Labor court and conclude that both employees and employers are roughly successful in their litigation.
Last, Ruhl et al.~\cite{Ruhl2017a} overviews some perspectives on how complex systems are useful for supporting policy-makers on legal-related topics such as appellate jurisprudence and tax policy analysis.

We are also motivated by recent results that show that intelligent systems can perform better than Law experts\footnote{\href{https://www.bbc.com/news/technology-41829534}{https://www.bbc.com/news/technology-41829534}}. Our hypothesis is that \emph{by taking advantage of Natural Language Processing (NLP) and Machine Learning techniques it is able to build a system that meets high quality legal decision predictions}. Different from the closest related works of this paper~\cite{Ashley2009,Aletras2016,Katz2017} which address United States and European courts, we propose an approach for legal decision prediction for Brazilian courts which also predicts whether the court decision will be unanimous. Moreover, in contrast to~\cite{Ashley2009,Aletras2016}, we trained a model at thousand-scale data set with $4,043$ cases.
Moreover, in contrast to~\cite{Aletras2016}, our approach does not rely on a binary classification problem -- since it uses three possible prediction results -- nor require that case data set should be separated by specific Law articles, hence being a more generic approach. 

The reminder of this paper is structured as follows. In Section~\ref{sec:methodology}, we present details on the aforementioned problem such as the case study and the methodology employed. Section~\ref{sec:results} exposes the results while Section~\ref{sec:conclusion} concludes our investigation and proposes future directions on this subject.

%
%
\section{Material and Methods}
\label{sec:methodology}

The research question which guides our study is \emph{how to predict legal decisions with a satisfactory level of accuracy for Brazilian courts by including the prediction of the court unanimous behavior}. Next sections provide further information about our assumptions and the proposed methodology.

\subsection{A generic approach}

We focus on Brazilian courts as Brazil legal system is not trivial. We believe that if we are able to solve this problem for such complex legal system, our approach would also fit other simpler or it could be straightforwardly adapted for more complex legal systems which share similarities. 
Nevertheless, it is worth to state that other legal systems also rely on similar documents. For instance, in Indiana Court of Appeal (United States), the Appellate Court decisions are composed by a group of three judges whose decisions (opinions) are divided in Case Summary, Facts, Procedural History, and the court conclusion at the end of that document.
In France, the Appellate Court (\emph{Cour d'appel}) also renders decisions coming from the agreement of three judges. That decision is called an \emph{Arr\^{e}t} whose structure is also composed of legal basis for the appeal, case history, and the final decision.

Further, we share the same assumption of Aletras et. al~\cite{Aletras2016}: ``there is enough similarity between (at least) certain chunks of the text of published judgments and applications lodged with the Court and/or briefs submitted by parties with respect to pending cases''.

\subsection{Decision labels and data set}
\label{sec:labels}

Regarding the flow process of a Brazilian appeal, when lawyers lodge appeals at a court it is analyzed by a group composed of three judges to check whether the appeal is able to be judged by the court.
If the appeal does not meet the formal requirements, the appeal is identified as not cognized (\emph{recurso n\~{a}o conhecido}) hence not judged by the court.
Otherwise, the appeal is therefore judged and classified in various categories. 
We therefore assumed that court decisions can be classified by using the following labels:
 
\begin{itemize}
    \item \texttt{not-cognized}, when the appeal was not accepted to be judged by the court;
    \item \texttt{yes}, for full favourable decisions;
    \item \texttt{partial}, for partially favourable decisions;
    \item \texttt{no}, when the appeal was denied;
    \item \texttt{prejudicada}, to mean that the case could not be judged for any impediment such as the appealer died or gave up on the case for instance;
    \item \texttt{administrative}, when the decision refers to a court administrative subject as conflict of competence between lower court judges.
\end{itemize}

In addition to the decision labels, an orthogonal concern of Brazilian court decisions -- as well as for other legal systems --  refers to its unanimity aspect, being labeled as:

\begin{itemize}
    \item \texttt{unanimity} which means that the decision was unanimous among the three judges that voted in the case; and
    \item \texttt{not-unanimity} by meaning that one of the judges disagreed on the  decision.
\end{itemize}

With respect to the data set, we relied on $4,762$ decisions (\emph{Ac\'ord\~aos}) from a State higher court (appellate court), the \emph{Tribunal de Justi\c{c}a de Alagoas}. 
From this data set, we removed the decisions that had repeated descriptions to not bias the sample thus resulting $4,332$ examples. 
Repeated decision descriptions occur owing to very similar cases which share the same description. 
Moreover, for the sake of predictability, we removed all the decisions classified as \texttt{prejudicada, not-cognized} and \texttt{admnistrative} as these labels refer to very peculiar situations which are not useful for prediction purposes addressed by this paper. Finally, the total amount of examples were $4,043$ cases.

\subsection{Methodology}

\begin{figure}[ht!]
    \centering
    \includegraphics[scale=.75]{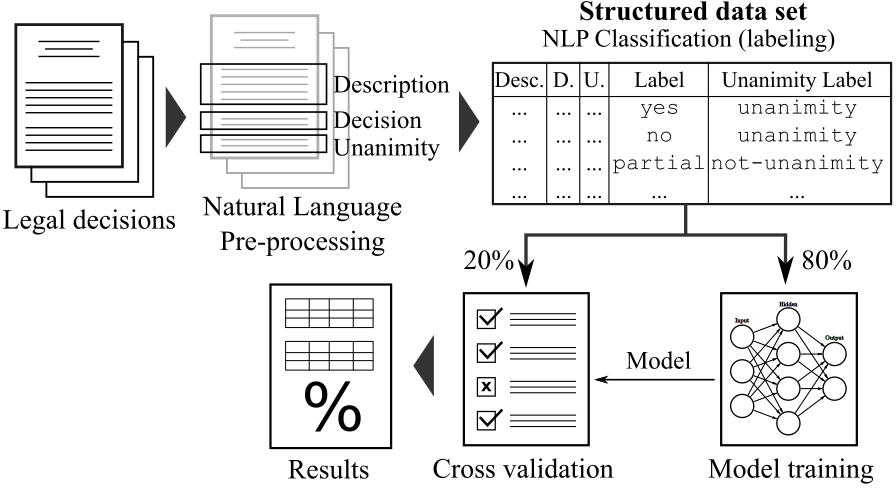}
    \caption{Methodology. Specific data from case outcomes are extracted then classified to different labels. A machine learning model is hence trained and assessed by using 5-fold cross-validation.}
    \label{fig:method}
\end{figure}

Figure~\ref{fig:method} depicts an overview of our methodology. 
From the legal decision data set, we extracted and separated the texts which hold information about the case description, the decision, and their unanimity aspect. This Natural Language Pre-processing task includes removing stop-words and word suffixes for improving the capacity of word representation. Then, we took advantage of Term Frequency-inverse Document Frequency (tdf-df) statistics to increase the importance of relevant words while decreasing the importance of general repetitive words not relevant to the addressed problem.
As follows, we used texts which refers to decisions and unanimity to classify them to one of the possible labels (c.f. Section~\ref{sec:labels}). As a result, we built a structured training data set depicted by Table~\ref{tab:dataset}.

\begin{table*}[ht]
  \centering
  \begin{tabular}{c|lll|ccp{2cm}}
    \textbf{Data} & \textbf{Decision description} & \textbf{Decision} & \textbf{Unanimity} & \textbf{Label} & \textbf{Unanimity label} \\ \hline
    \textbf{Sample 1} & Direito Processual Civil... &  Recurso conhecido e provido &  Unanimidade & \texttt{yes} & \texttt{unanimity} \\ \hline 
    \textbf{Sample 2} & Apela\c{c}\~{a}o criminal... &  Recurso conhecido e parcialmente provido &  Decis\~{a}o unânime & \texttt{patial} & \texttt{unanimity} \\ \hline
    \textbf{Sample 3} & Apela\c{c}\~{a}o C\'{i}vel em A\c{c}\~{a}o Ordin\'{a}ria... &  Recurso conhecido e n\~{a}o provido & Decis\~{a}o un\^{a}nime & \texttt{no} & \texttt{unanimity} \\ \hline
  \end{tabular}
   \vspace*{0.5mm}
    \caption{Training data set includes decision texts and labels which were classified according to respective decision texts. E.g., in Sample 1, \emph{provido} was classified as a favorable (\texttt{yes}) decision and \emph{Unanimidade} was classified as \texttt{unanimity}.}
    \label{tab:dataset}
\end{table*}

Next, we used $80\%$ of the data set to train a Machine Learning model which was then assessed by using the latter $20\%$ of the data set (c.f. Figure~\ref{fig:method}). 
To train a model means to automatically find out which parameters are the most suitable for predicting decisions based on the training data set.
Because we address decision and unanimous predictions, it requires to train two models to address both predictions. Once trained, the models can be used to predict decision and unanimity given a case description. 
Last, to evaluate our approach, we used the F1-score metrics and performed $5$-fold cross-validation to improve the practicability of our approach. Results are thus exposed as success accuracy rate in percentage.

Furthermore, in order to assess the exposed methodology, we developed a working prototype in Python.
We used the NLTK framework~\cite{Loper2002} for Natural Language Processing in such a way that our prototype is easily configurable for various languages in addition to Portuguese. 
The prototype also provides a graphical user interface which can be accessible from any Web browser.

%
%
\section{Results}
\label{sec:results}

Our approach was able to score $\textbf{78,99\%}$ F1-score ($\sigma^2=0.000017$) when predicting legal decision for the \emph{Tribunal de Justi\c{c}a de Alagoas} Brazilian court by using $4,043$ judge decisions.
The number of samples for each label is exposed in Table~\ref{tab:decision-dist}.

\begin{table}[!ht]
\begin{center}
\begin{tabular}{lllll}
\hline
\textbf{Labels} & \texttt{no} & \texttt{partial} & \texttt{yes} \\ \hline
\textbf{N. of decisions} & 2,415 & 866 & 762 \\ \hline 
\end{tabular}
\end{center}
\caption{Distribution of decisions according to their labels.}
\label{tab:decision-dist}
\end{table}

In order to analyze our prediction over a more uniformly distributed data set, we randomly removed $1,549$ \texttt{no}-labeled decisions to have the same number of \texttt{partial}-labeled decisions. Table~\ref{tab:decision-dist-shorter-dataset} depicts the distributions of each decision label. The accuracy of case outcome prediction in this situation was $\textbf{74.07\%}$ ($\sigma^2=0.00029$) for the F1-score metrics. This result is very interesting as it does not validate our previous assumption that the not regularly distributed data set would strongly bias the model.

\begin{table}[!ht]
\begin{center}
\begin{tabular}{lllll}
\hline
\textbf{Labels} & \texttt{no} & \texttt{partial} & \texttt{yes} \\ \hline
\textbf{N. of decisions} & 866 & 866 & 762 \\ \hline 
\end{tabular}
\end{center}
\caption{Distribution of decisions according to their labels when randomly removing \texttt{no}-labeled decision samples to create a regular distributed data set.}
\label{tab:decision-dist-shorter-dataset}
\end{table}

With respect to predicting the unanimous behaviour of the \emph{Tribunal de Justi\c{c}a de Alagoas} Brazilian court, our approach scored $\textbf{98.46\%}$ ($\sigma^2=0.000031$) for the F1-score metrics. This assessment was performed in a data set with $2,274$ cases. From the $4,332$ data set -- which had no repeated decision descriptions --, we removed the samples that either our classifier did not managed to label or the decision itself did not had any information about unanimity. The resulting data set had $2,289$ samples. As follows, we removed from this data set the decision whose labels were \texttt{prejudicada, not-cognized} and \texttt{admnistrative} -- as these labels are not relevant for the predictive addressed problem -- resulting in a data set with $2,274$ examples. The distribution of \texttt{unanimity} and \texttt{not-unanimity} labels are depicted by Table~\ref{tab:unanimity-dist}.

\begin{table}[!ht]
\begin{center}
\begin{tabular}{lllll}
\hline
\textbf{Labels} & \texttt{not-unanimity} & \texttt{unanimity} \\ \hline
\textbf{N. of decisions} & 45 & 2,229  \\ \hline 
\end{tabular}
\end{center}
\caption{Distribution of decisions according to judge unanimous behavior.}
\label{tab:unanimity-dist}
\end{table}

The very-high unanimous predictive accuracy of $98.46\%$ is explained by the fact that most of decisions are unanimous, therefore the model was biased to this label. We indeed expected that most decisions were unanimous since this is well known by law experts. However, the great difference between unanimous and non unanimous decision is a surprising result.
In order to understand how our approach would perform when predicting unanimity by using a more uniformly distributed data set, we therefore performed another evaluation by randomly removing decisions whose label was \texttt{unanimity} to meet the same number of \texttt{not-unanimity} decisions. The resulting data set had $90$ examples, half of them labeled as \texttt{unanimity} and the other half \texttt{not-unanimity}. With this configuration, our prototype reached $\textbf{76.94\%}$ ($\sigma^2=0.015$) F1-score accuracy.

%
%
\section{Conclusion}
\label{sec:conclusion}

This paper proposes a methodology for predicting Brazilian court legal decisions which is able to reach $79\%$ of accuracy when employed for a Brazilian court data set with $4,043$ cases. 
Our approach is able to predict case outcomes by using three different labels: \emph{yes, no, partial}.
Moreover, the proposed method also predicts whether the decision will be unanimous, which fits not only Brazil legal system, but also several others whose decisions are judged by more than one judge. The unanimity prediction performance of our approach is $77\%$ of accuracy. To our knowledge, this is the first study to predict Brazil legal decisions.

Moreover, our approach is easy to use as it only requires that users provide the description of their litigation and the output will be one of the aforementioned case outcome predictive label along with its predictive unanimity label.
These information are very useful for attorneys, judges, and other Law professionals as they provide practical support for their work.
Moreover, our contribution also includes a working prototype which can be configured to further languages as well as for different data sets. 

Although we believe that our contribution is quite satisfactory given the accuracy rate aforementioned, future investigations might consider comparing our results with Law experts, as performed in~\cite{Ruger2004} and by current Lawtech products such as Case Crunch and LawGeex\footnote{\href{https://www.artificiallawyer.com/2018/02/26/lawgeex-hits-94-accuracy-in-nda-review-vs-85-for-human-lawyers/}{https://www.artificiallawyer.com/2018/02/26/lawgeex-hits-94-accuracy-in-nda-review-vs-85-for-human-lawyers/}}.
Other future work includes to investigate whether taking advantage of existent Named-entity recognition data sets for Brazilian law documents~\cite{Araujo2018} improve the prediction quality. 
Furthermore, the assessment of the proposed method can be performed on larger and/or different data sets, such as the European Court of Human Rights for instance. Ultimately, despite the various directions one might take to leverage our work, we believe that Mireille Hildebrandt's viewpoint on ``agnostic machine learning'' and its consequences to the Rule of Law~\cite{Hildebrandt2018} should be taken into account when designing and using a legal predictive system.


\end{document}